\documentclass{elsart}
\usepackage{graphicx,amssymb}

\begin{document}
\begin{frontmatter}
\title{Experimental Lagrangian Acceleration Probability Density Function Measurement}
\author{N. Mordant, A.~M. Crawford and E. Bodenschatz}
\address{Laboratory of Atomic and Solid State Physics, Cornell University, 14853 Ithaca NY}

\begin{abstract}
We report experimental results on the acceleration component probability distribution function at $R_\lambda = 690$ to probabilities of less than
$10^{-7}$.
This is an improvement of more than an order of magnitude over past measurements
and allows us to conclude that the fourth moment
converges and the flatness is approximately 55.
We compare our probability distribution to those predicted by several models inspired by non-extensive statistical mechanics.
We also look at acceleration component probability distributions conditioned on a  velocity component for conditioning velocities as
high as 3 times the standard deviation and find them to be highly non-Gaussian.

\end{abstract}
\end{frontmatter}
\maketitle

\section{Introduction}

Recent developments of new experimental techniques have allowed significant progress in the
Lagrangian description of turbulence~\cite{jfm,rsi,nature,lyon,lyon1,riso}. Although this aspect has been investigated
theoretically since the beginning of the 20th century, the extreme difficulty of recording particle trajectories in high
Reynolds number turbulence restricted experimental investigations. The availability of very high speed imaging sensors gives access to the Lagrangian acceleration in fully developed turbulence. It has been shown that the probability density function (PDF) of a Lagrangian acceleration component is strongly non-Gaussian, {\it i.e.}, the PDF exhibits extremely large tails~\cite{jfm,nature}. This has been also observed in direct numerical simulations of fully developed turbulence~\cite{vedula,gotoh}

These observations resemble features obtained in the field of non-extensive statistical mechanics~\cite{beck,arimitsu,reynolds,becklog}.
This opens two promising perspectives: first toward a statistical theory of turbulence and second toward a dynamical modeling of the motion of fluid particles.
The latter comes from the theoretical link first established by C. Beck~\cite{beck} between the Tsallis entropy and a stochastic equation for the Lagrangian acceleration.
These developments led to a series of models for the PDF of Lagrangian accelerations. The multitude of  models needs to be tested by an accurate comparison
with experimental data, especially in the low probability tails of the acceleration PDF.

Here we present new experimental results that have allowed us to improve the estimation of the acceleration PDF.
First we describe the modifications that have been implemented on the previous experiment, then we show the acceleration PDF and  compare it to
the different classes of models.
Finally we show and discuss some PDF's of acceleration conditioned on velocity which are improvements over results presented previously in Sawford et al.~\cite{Saw:2003}.

\section{The experiment}

The experimental set-up has been described in detail in previous publications~\cite{jfm,rsi,nature}. The reader should refer to these
for detailed information on the experiment.
Let us recall briefly the main features.
The flow under consideration is of the Von~K\'arm\'an kind: the water is driven by two counter-rotating disks, 20~cm in diameter, 33~cm apart,
mounted in a cylindrical container.
The rotation frequency of the disks is regulated at 3.5~Hz for the experiment under consideration here.
The integral scale of the flow has been shown to be 7~cm and the energy dissipation rate to be 1.14~m$^{2}$s$^{-3}$~\cite{jfm}. The transverse and axial spatial coordinates of small tracer particles (diameter 46~$\mu$m, density $\rho = 1.06 g/cm^3$) are recorded by two silicon strip detectors. Each detector provides a 1D recording of the particle trajectories at a sampling rate up to 70~kHz. The root-mean-square velocity is 0.43~ms$^{-1}$, so that the Taylor based Reynolds number is 690.
The Kolmogorov time  is $\tau_{\eta}=0.93$~ms, the sampling frequency is $F_{s}=69.9$~kHz so that $\tau_{\eta}$ corresponds to 65 samples. We study here only one transverse component of the acceleration. Even though the velocity components have been shown to be anisotropic, the acceleration components are close to isotropic at that Reynolds number~\cite{jfm}.

Compared to the experiment reported earlier, we are now able to record  data three time faster (in terms of global efficiency,
{\it i.e.},  including transfer rate from the digitizer to the computer, repetition rate, etc.) because of  a hardware upgrade of the data-acquisition computers. This allowed us  to
obtain more data in a reasonable time. The laser used for illumination has also been upgraded to a  frequency doubled pulsed YAG laser with  mean output power
as high as  35 watts, a typical  pulse duration of  200~ns and  a pulse repetition rate of up to 70~kHz. This pulsed laser improves the
efficiency of the silicon strip detectors due to the short duration of the pulses.

Another difference lies in the signal processing. As stressed in  Voth et al. \cite{jfm}, one has to low-pass filter the trajectories to filter the noise in order to have access to the acceleration. In our previous work  \cite{jfm,nature}, the filtering and differentiation were achieved by fitting a parabola to the data, in a time window whose duration had to be appropriately chosen.
Here, this procedure has been modified to improve the performance of the data processing algorithm. We now use a differentiating and filtering kernel $k(\tau)$ so that the acceleration is obtained by convolution
$a=k\star x$, where $\star$ is the convolution product and $x$ is the spatial coordinate of the particle.
 We have chosen a Gaussian kernel
\begin{equation}
g(\tau)=\frac{1}{\sqrt{\pi w^2}} \exp\left(-\frac{\tau^2}{w^2}\right) \, .
\end{equation}

To achieve both a  low pass filter and the second order differentiation of trajectories, one simply has to differentiate twice the Gaussian kernel
\begin{equation}
k(\tau)=\frac{2}{\sqrt{\pi} w^3}\left(\frac{2\tau^{2}}{w^{2}}-1\right) \exp\left(-\frac{\tau^2}{w^2}\right)\, .
\end{equation}
In practice, however,  time is discrete and one has to use a finite time window, therefore the kernel is truncated. This requires a modification of the normalization. The following kernel was  used in the data analysis (for discrete times):
\begin{eqnarray}
k(\tau)&=&A\left(\frac{2\tau^{2}}{w^{2}}-1\right) \exp\left(-\frac{\tau^2}{w^2}\right)+B\;\;\textrm{ if } | \tau | \leqslant L\\
k(\tau)&=&0 \;\;\textrm{ otherwise}
\end{eqnarray}
where $A$ and $B$ are normalization constants chosen to ensure that $k\star1=0$ and  $k\star\tau^{2}=2$,  {\it i.e.}, the second derivative of a constant is 0 and
the second derivative of a parabola is 2. Note that, if $L$ equals one time sample $1/F_{s}$, one recovers the simple 3 points discrete second
derivative (independently of the value of $w$). If $L\gg w \gg 1/F_{s}$ one converges to the continuous kernel $g$ when $F_{s}$ goes to infinity.
The results obtained from the filtering operation are found to be in agreement with  those obtained by the parabolic fitting procedure.
Note that due to the truncation, the actual width of the filter is a slightly lower that $w$.

\section{Probability density function of Lagrangian acceleration}

\subsection{Experimental estimation of the PDF}

\begin{figure}[htbp]
\begin{center}
\includegraphics[width=\textwidth]{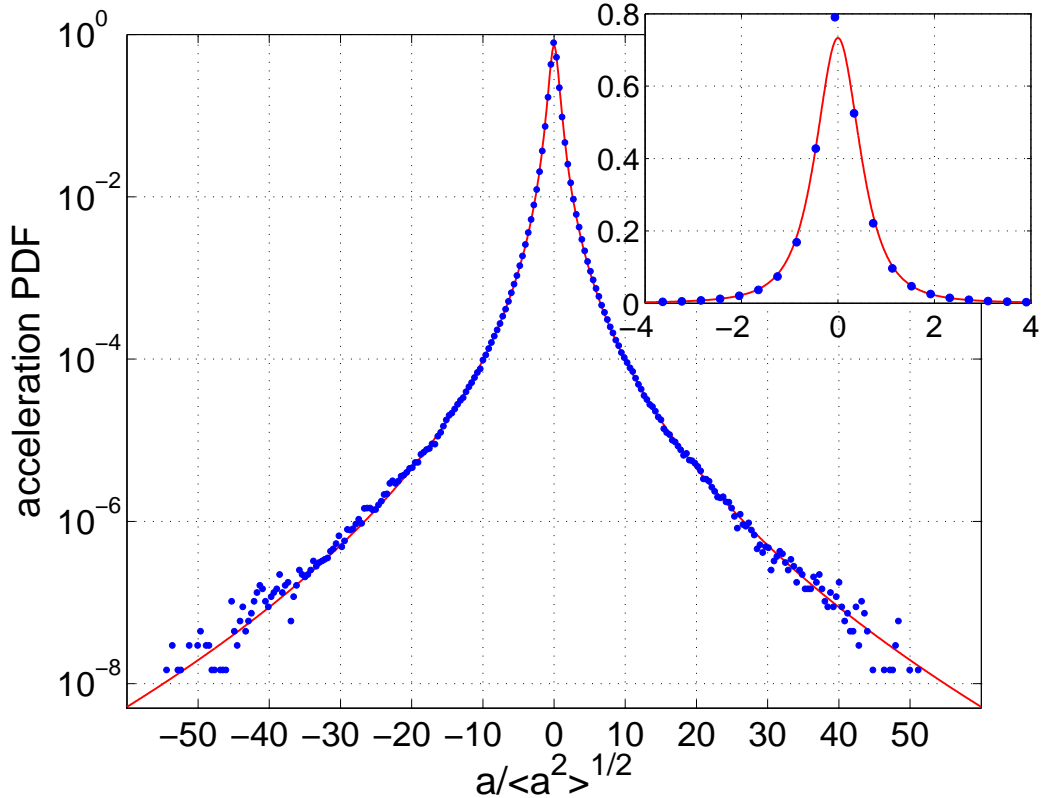}
\caption{Lagrangian acceleration PDF at $R_{\lambda}=690$. The kernel parameters are $w=16$ samples, $L=1.5w$. The solid line is the stretched exponential fit. Insert: linear plot.}
\label{pdf}
\end{center}
\end{figure}
The results presented here required continuous acquisition of data for two weeks.
The data set consists of $1.7\; 10^{8}$ points, which include the new measurements and the data reported in the previous publications \cite{jfm,nature}.
The experimental probability density function (PDF) is presented in Fig.~\ref{pdf} as dots.
The new data allows the estimation of the PDF to reach probabilities more than one order of magnitude lower than the results in~\cite{jfm,nature}, {\it  i.e.},  less than $10^{-7}$. Fig.~\ref{pdf} also shows the stretched exponential fit:
\begin{equation}
P(a)=C\exp\left(-a^2 / \left(1+\left|\frac{a\beta}{\sigma}\right|^\gamma\right)\sigma^2 \right)
\end{equation}
with $\beta=0.513\pm0.003$, $\gamma=1.600\pm0.003$, $\sigma=0.563\pm 0.02$ and $C$ is a normalization constant ($C=0.733$). This expression is the same as the one proposed earlier ~\cite{jfm,nature}, with slightly different values due to improved accuracy of the fit and perhaps also a difference in the Reynolds number. The initial values were $\beta=0.539$, $\gamma=1.588$, $\sigma=0.508$ at $R_{\lambda}=970$. This stretched exponential fit is shown to follow the experimental  data  very well.

\begin{figure}[htbp]
\begin{center}
\includegraphics[width=\textwidth]{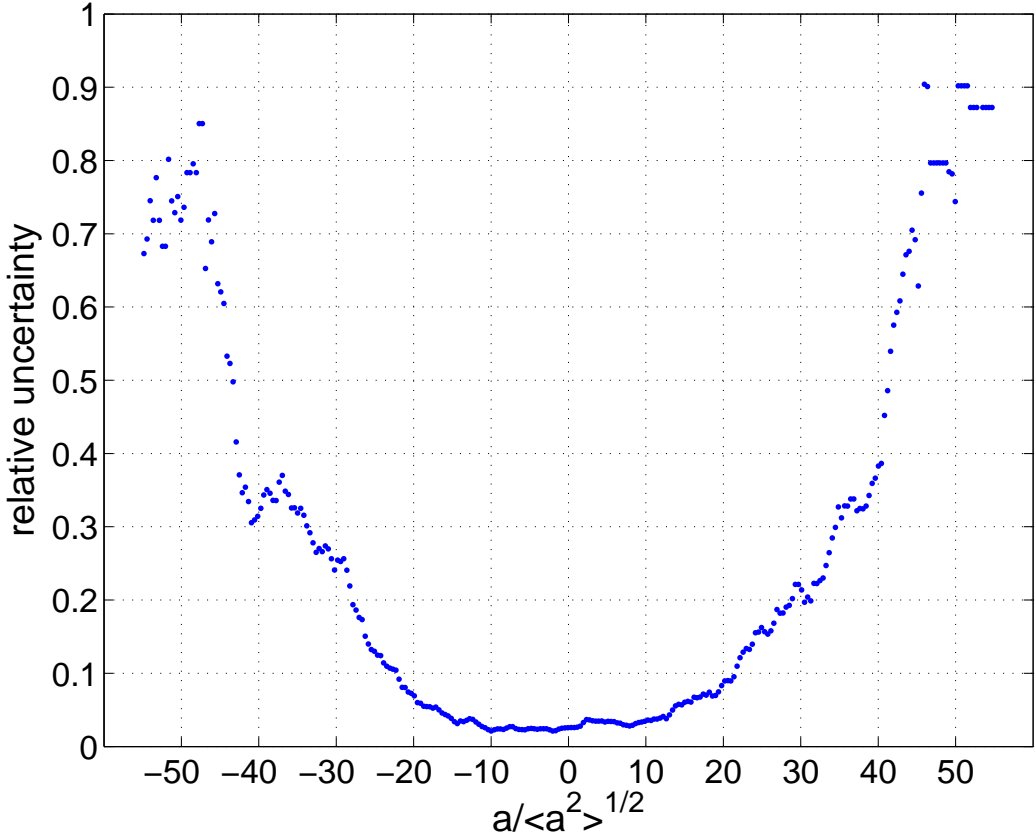}
\caption{Relative uncertainty on the Lagrangian acceleration PDF at $R_{\lambda}=690$. The {\it rms} error is estimated by splitting the data set in 6 subsets.}
\label{error}
\end{center}
\end{figure}
In Fig.~\ref{error}, an estimation of the statistical dispersion of the PDF is shown. It has been obtained by splitting the whole data set into 6 subsets. The relative {\it rms} uncertainty is displayed. One observes a minimum of 3\% for small values of the acceleration, $|a|<10 \langle a^{2}\rangle^{1/2}$, for which the number of samples is maximum. The uncertainty then increases as $|a|$ increases. It remains lower than 40\%  for $|a|<40\langle a^{2}\rangle^{1/2}$.

\begin{figure}[htbp]
\begin{center}
\includegraphics[width=\textwidth]{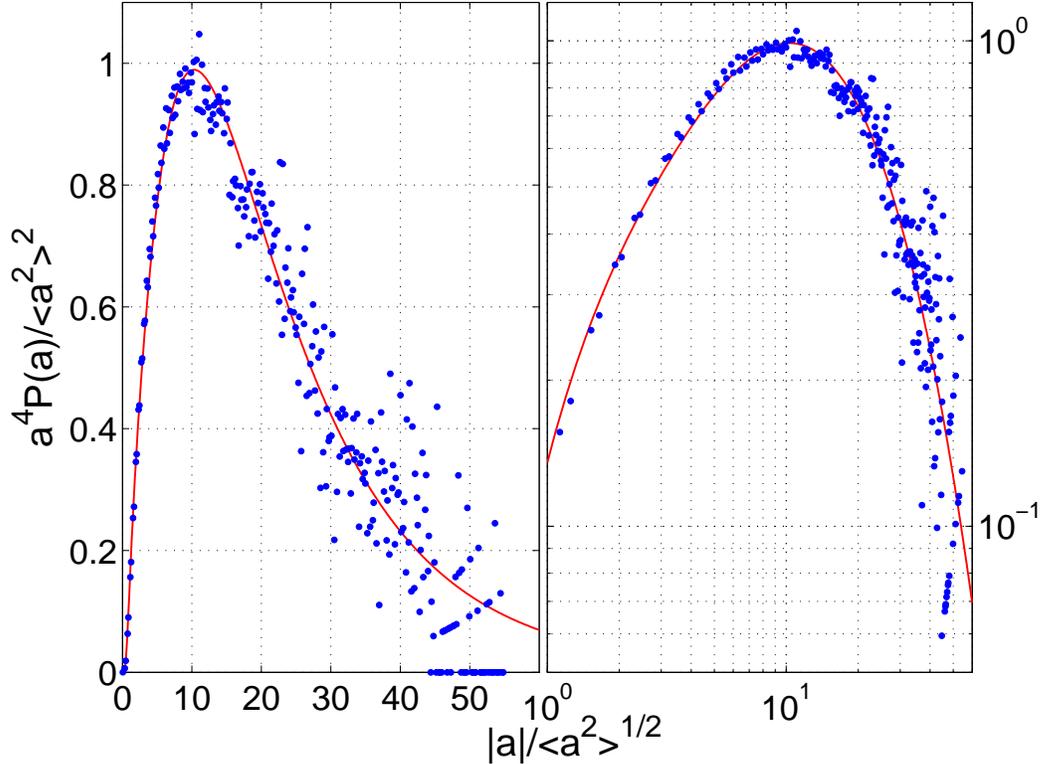}
\caption{Lagrangian acceleration PDF at $R_{\lambda}=690$. The kernel parameters are $w=16$ samples, $L=1.5w$. The solid line is the stretched exponential fit.}
\label{flatness}
\end{center}
\end{figure}
One consequence of our better estimation of the acceleration PDF is that we can conclude that the fourth moment converges.
Fig.~\ref{flatness} displays the contribution to the fourth moment: $a^{4}P(a)$. The first observation is that this function decreases
as $|a|$ increases. From our fit, the decay is consistent with a stretched exponential:
\begin{equation}
P(a)\propto\exp (-|a|^{0.4}) \, .
\end{equation}
This means that the integral $\int a^{4}P(a)$ converges. The data gives  a flatness of
\begin{equation}
F=\frac{\langle a^4\rangle}{\langle a^2\rangle^2}=55\pm8 \, .
\end{equation}
and the  stretched exponential fit gives a flatness value equal to $F=55.2$. The possible influence of the particle response time on this value is expected to be low as the characteristic time of the particle is $0.12\tau_\eta$. The filtering procedure has a limited impact because of the low value of the filter width. Changing the width from $0.23\tau_\eta$ to $0.31\tau_\eta$ decreases the variance by 10$\%$ and the flatness by 15$\%$. The shape of the PDF is unchanged for $a/\langle a^2 \rangle^{1/2}<25$. Thus we expect the flatness to be slightly underestimated but do not expect any qualitative change in the shape of the PDF.

\subsection{Comparison with models inspired by non-extensive statistical mechanics }

Among the large class of superstatistics~\cite{beckcohen}, four specific examples have been proposed to model the observed acceleration PDF. The first one, proposed by C.~Beck~\cite{beck} (and referred to as Beck-$\chi^2$ here after), is based on the $\chi^2$-distribution. One very interesting feature of this analysis is the possibility of making a connection between a statistical approach and a dynamical approach. It allows us to build a stochastic equation for the Lagrangian time dynamics of acceleration. In the same spirit, A.~M.~Reynolds~\cite{reynolds} and then C.~Beck~\cite{becklog} proposed a second model (referred as Reynolds and Beck-Log resp.) based on log-normal statistics. The approach of Reynolds is much more ambitious. He tries to build a full stochastic equation that models the acceleration. This model takes into account not only the small time scales dynamics but also the large scales. He takes all numerical parameters from experiments or direct numerical simulations reported in the literature. Therefore there are no free parameters in his model, once the Reynolds number is chosen. Thus the result of the model is not a fit to the experimental data.

Another more classical approach  has been proposed by T.~Arimitsu and N. Arimitsu~\cite{arimitsu} (referred as Arimitsu).
They use the properties of scale invariance of the Navier-Stokes equation to build a multifractal description of the acceleration PDF.

\begin{figure}[htbp]
\begin{center}
\includegraphics[width=\textwidth]{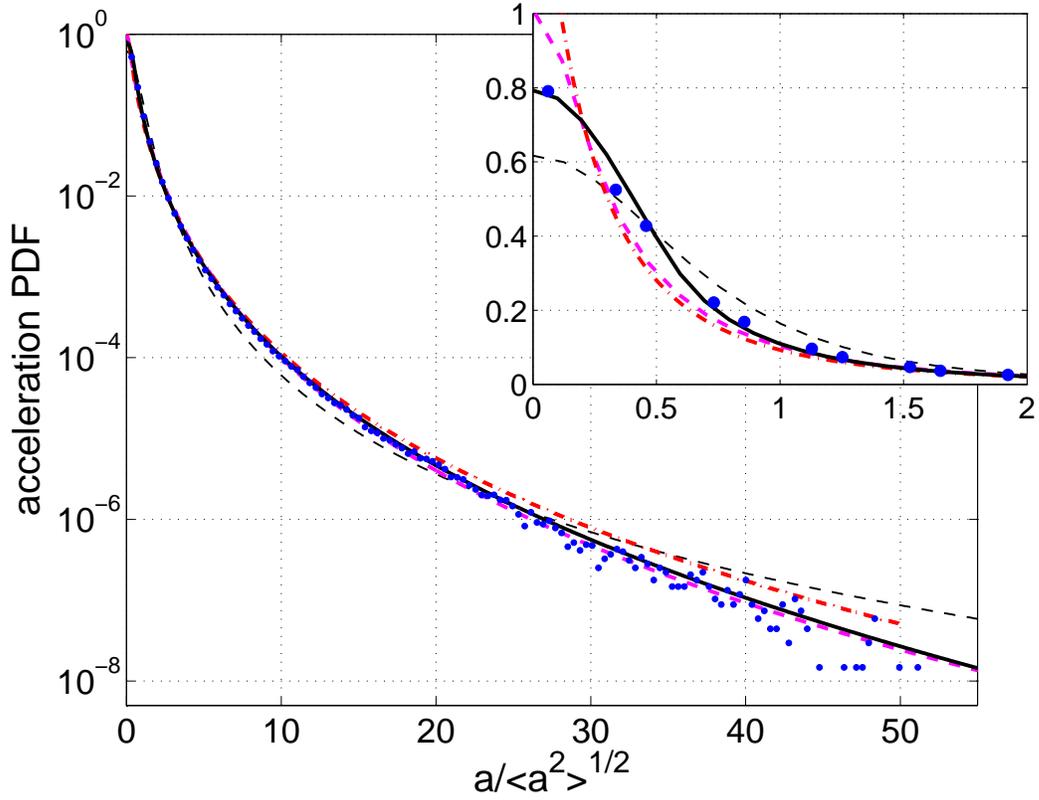}
\caption{Comparison of the experimental acceleration PDF with the models. Dots: data, thin dashed line: Beck-$\chi^2$, $q=\frac{3}{2}$~\cite{beck}. Thick dashed line: Beck-Log, $s^2=3.0$~\cite{becklog}. Solid line: Arimitsu, $\mu=0.25$ and $n=17.1$~\cite{arimitsu}. Dot-dashed: Reynolds~\cite{reynolds}. Insert: linear plot.}
\label{model}
\end{center}
\end{figure}

\begin{figure}[htbp]
\begin{center}
\includegraphics[width=\textwidth]{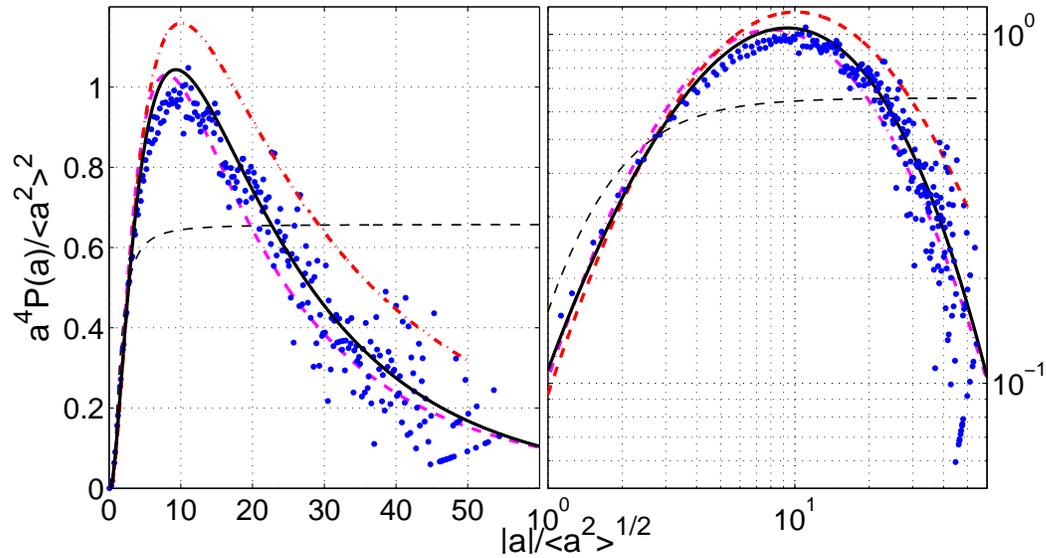}
\caption{Contributions to the fourth order moment $a^4P(a)$. Left: linear plot, right: loglog plot. Same symbols as previous figure.}
\label{modelflat}
\end{center}
\end{figure}

These four models are displayed in figures~\ref{model} and~\ref{modelflat} together with the experimental data. The first obvious observation is that
the $\chi^2$-distribution model (with $q=3/2$) does not fit the data. The decay of the tails displays  qualitatively incorrect behavior.
Indeed, for any value of the parameter $q$, the model predicts a power-law decay which is not observed. For $q$ close to 3/2, which gives the best fit,
the fourth moment diverges which is not the case for the experimental data.
The modification of the distribution to log-normal, introduced later by C. Beck, makes the fit qualitatively much better.
In particular the flatness, for $s^2=3.0$ takes a finite value $F=60.3$.
The Reynolds model does not agree as well with the data as the others. One reason is that there are no free parameters in this model. Second, we low-pass filter the acceleration at the scale $0.23\tau_\eta$. This scale is small but not zero. 
If the temporal position signals from the Reynolds model are low-pass filtered at the same scale then the overlap with the experimental data is highly improved~\cite{reynolds1}.
The multifractal model by Arimitsu also provides a good agreement with the experimental results.
Concerning the region of low acceleration region, the Arimitsu model captures the correct behavior, contrary to the other models. Reynolds' and Beck's log-normal models overestimate the zero acceleration probability whereas the Tsallis statistics model underestimates it.

For all these models the departure from the experimental curve for large accelerations is best viewed in Fig.~\ref{modelflat}.
Most of the proposed models do not peak at the same value of acceleration as the experimental curve. Nevertheless most of them are very close to the experiment.

\subsection{Conditional probabilities}

\begin{figure}[htbp]
\begin{center}
\includegraphics[width=\textwidth]{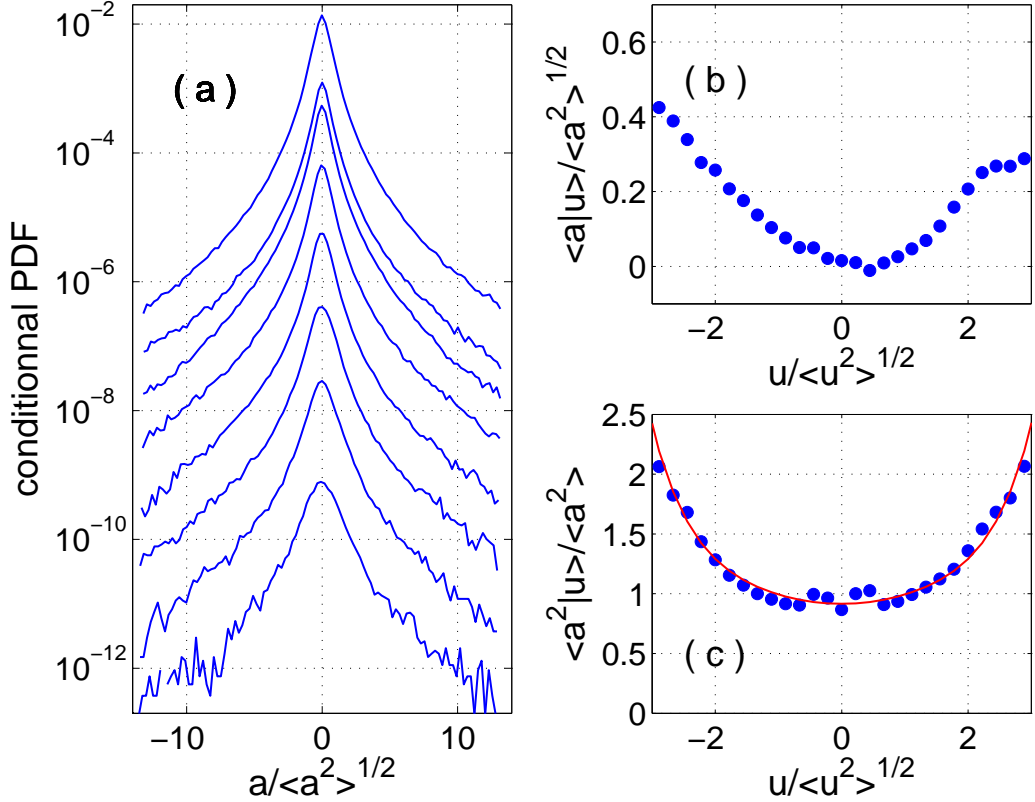}
\caption{(a) Acceleration PDF's conditional on velocity.
From top to bottom, condition: $u= 0, 0.45, 0.89, 1.3, 1.8, 2.2, 2.7\textrm{ and }3.1\,u_\mathrm{rms}$.
The curves have been shifted for clarity by a factor 2, the top one being unchanged. (b) The conditional mean acceleration. (c) The conditional acceleration variance. Dots are the experiment, the solid line is a 6th order polynomial fit.}
\label{cond}
\end{center}
\end{figure}
Conditional acceleration statistics are important in  many second order 
stochastic models~\cite{reynolds,Saw:2003,Saw:1991,Saw:1998,borgas:1998}.
A  very large number of statistics is required to obtain acceleration PDF's conditional 
on large velocities since large velocities are rare events.
The new large data set allowed us to calculate distributions conditional on larger velocities 
than those shown in Sawford et. al.~\cite{Saw:2003}.
Figure~\ref{cond}  shows PDF's of the transverse component of acceleration 
conditional on the transverse component of velocity.
The conditional probability distributions are highly non-Gaussian. They have almost the same 
shape as the unconditional distribution.
The conditional acceleration variance increases with increasing velocity which seems to be 
due mainly to a widening of the tip of the
distribution as can be see in in Figure~\ref{cond}. For homogeneous isotropic turbulence the 
conditional mean acceleration should
be zero. Departures from zero  reflect the anisotropy of our flow 
although DNS of homogeneous isotropic turbulence has
also shown slight departures from zero~\cite{Saw:2003}. 
Sawford et. al. present a scaling argument that indicates that
the conditional acceleration component variance $\langle A_i^2 | u_i \rangle \approx u_i^6$ to leading order. A fit to our data in Fig.~\ref{cond}c. shows good agreement with this.
The dependence of acceleration distributions on the velocity violates local homogeneity~\cite{Pope:book}
which is an assumption of Kolmogorov 41 theory.

\section{Conclusion}

Our new set of data helps to discriminate among different classes of models. In particular, the model based on the Tsallis entropy is shown to disagree with the experimental observations: the acceleration PDF does not display any power law decrease and is well fitted by a stretched exponential decay. One consequence is that the fourth moment of the acceleration component is finite and we observe a value of the flatness close to 55. The models based on log-normal statistics proposed by Beck and Reynolds or multifractal analysis (Arimitsu) are indeed close to our experimental observations. Since the only difference between Beck's two models is the underlying statistics (Tsallis or log-normal), the hypothesis of Tsallis statistics must be discarded in order to reproduce the observed behavior of the acceleration PDF. The use of a log-normal statistics and the physical link of the $\beta$ parameter in Beck's models to the dissipation rate are in agreement with the refined 1962 Kolmogorov-Obukhov theory of turbulence which is known to describe well the features of intermittency.

We also show that the joint PDF of acceleration and velocity is not gaussian. The acceleration variance conditioned on velocity is also not constant. These results have important consequences on the structure of the stochastic equation required to model dispersion by turbulent flows. 

\begin{ack}
This work has been supported by NSF grant \#9988755. We thank Arthur La~Porta for valuable discussions and writing substantial parts of the data analysis code.
\end{ack}

 \end{document}